\documentclass[journal]{IEEEtran}

\usepackage{blindtext}
\usepackage{makecell,multirow,diagbox}
\usepackage{subfigure}
\usepackage{enumerate}
\usepackage{url}
\usepackage{booktabs}
\usepackage{color}
\usepackage{graphicx}
\usepackage{amsmath, amssymb}

\ifCLASSINFOpdf
\else
\fi
\begin{document}
%
\title{Inferring Multiple Relationships between ASes using Graph Convolutional Network}
%
%
%

\author{Songtao~Peng,
        Xincheng~Shu,~\IEEEmembership{Student Member,~IEEE,}
        Zhongyuan~Ruan,
        Zegang~Huang,
        and~Qi~Xuan,~\IEEEmembership{Member,~IEEE}
\thanks{S. Peng, X. Shu, Z. Ruan, Z. Huang and Q. Xuan are with the Institute of Cyberspace Security, College of Information Engineering, Zhejiang University of Technology, Hangzhou 310023, China. e-mail: pengst.aiing@gmail.com; sxc.shuxincheng@foxmail.com; zyruan@zjut.edu.cn; gre35975@gmail.com; xuanqi@zjut.edu.cn.}
\thanks{Corresponding author: Qi Xuan.}}

%
%

\markboth{}%
{Shell \MakeLowercase{\textit{et al.}}: Bare Demo of IEEEtran.cls for IEEE Journals}
%



\maketitle

\begin{abstract}
Precisely understanding the business relationships between Autonomous Systems (ASes) is essential for studying the Internet structure. So far, many inference algorithms have been proposed to classify the AS relationships, which mainly focus on Peer-Peer (P2P) and Provider-Customer (P2C) binary classification and achieved excellent results. However, there are other types of AS relationships in actual scenarios, i.e., the business-based sibling and structure-based exchange relationships, that were neglected in the previous research. These relationships are usually difficult to be inferred by existing algorithms because there is no discrimination on the designed features compared to the P2P or P2C relationships.

In this paper, we focus on the multi-classification of AS relationships for the first time. We first summarize the differences between AS relationships under the structural and attribute features, and the reasons why multiple relationships are difficult to be inferred. We then introduce new features and propose a Graph Convolutional Network (GCN) framework, AS-GCN, to solve this multi-classification problem under complex scene. The framework takes into account the global network structure and local link features concurrently. The experiments on real Internet topological data validate the effectiveness of our method, i.e., AS-GCN achieves comparable results on the easy binary classification task, and outperforms a series of baselines on the more difficult multi-classification task, with the overall accuracy above 95\%.
\end{abstract}

\begin{IEEEkeywords}
Autonomous systems, multiple relationships, graph convolutional network, inference algorithm, BGP topology.
\end{IEEEkeywords}

%
\IEEEpeerreviewmaketitle

\section{Introduction}
%
%
%
%
\IEEEPARstart{A}{s} a typical complex network, the Internet is now composed of more than 70000 Autonomous Systems (ASes), where an AS is a network unit that has the right to independently decide which routing protocol should be used in the system. Routing information is then exchanged between ASes through Border Gateway Protocol (BGP) to achieve global reachability. Meanwhile, routing information can be effectively used to construct the topological graph of the AS interconnection and thus can implement strategic decisions at the AS-level. Typically, in AS-level topology, the business relationships between connected ASes are broadly classified into (1) Customer-Provider (C2P), (2) Peer-Peer (P2P), and (3) Sibling relationship (S2S). A series of studies on these relationships include topology flattening~\cite{gill2008flattening, labovitz2010internet}, network congestion detection~\cite{sundaresan2017challenges, dhamdhere2018inferring, smith2018routing}, Internet security check~\cite{cho2019bgp, cohen2016jumpstarting, gill2011let, karlin2008autonomous}, variable routing protocol based attack designing~\cite{apostolaki2018sabre}, and network characteristics analysis~\cite{tozal2018policy}. 

In this paper, we focus on the accurate knowledge of AS business relationships, which is essential for understanding both technical and economic aspects of the Internet. Since the AS relationship was defined in 2001~\cite{gao2001inferring}, many inference algorithms~\cite{dimitropoulos2007relationships,gregori2011bgp,luckie2013relationships,susan2015machine,jin2019stable,jin2020toposcope,shapira2020unveiling, giotsas2014inferring} have been proposed and proved the superiority of their performance. These well-recognized researches share some common characteristics: (1) BGP data come from some operators make BGP routing information from their routers available, which are used for monitoring, troubleshooting and research purposes. (2) Internet Exchange Point (IXP) and sibling relationship are obtained from the PeeringDB dataset~\cite{peeringdb} and AS-to-organization mapping data~\cite{as2org}, and community characteristics are concerned~\cite{giotsas2012valley}. (3) Their focus is shifted from most easily inferred links to a small number of critical links that are difficult to infer. Notably, it is found that some advanced algorithms, including AS-Rank~\cite{luckie2013relationships}, ProbLink~\cite{jin2019stable}, TopoScope~\cite{jin2020toposcope}, have a lower prediction error rate~\cite{luckie2013relationships}, and behave surprisingly well in predicting \textit{hard links}~\cite{jin2019stable} or \textit{hidden links}~\cite{jin2020toposcope}. However, most of the current experimental environments simplifies the Internet network, and the above studies just infer P2P and P2C without considering multiple relationships but simply let sibling relationship as known.

In order to explore the real situation, we pay attention to those complex relationships that are small in number but essential in the Internet. As a centralized exchange platform established between different operators to connect their respective networks, the IXP is an indispensable part of the high-level Internet structure. We have carefully considered the structure-based Exchange (X2X) relationship and the three traditional business relationships. By synthesizing and analyzing the important indicators in each advanced inference method, like \emph{triplet}, \emph{distance to clique} and so on~\cite{luckie2013relationships}, we find that the P2C and S2S relationships have quite high similarities, making it hard to infer the relationship between ASes of the same organization. Meanwhile, the features of X2X relationship are also very similar to those of P2P. In other words, it is quite difficult to distinguish S2S from P2C, and X2X from P2P, while it is relatively easily to classify P2C and P2P, in the current inference framework.

To solve this multi-classification relationship inference problem which is a more challenging, we introduce two new features, \emph{common neighbor ratio} and \emph{AS type}, and develop a new Graph Convolutional Network (GCN) framework AS-GCN. The convolution process of AS-GCN on the Internet topology is a good simulation of the \emph{valley-free} path characteristic~\cite{gao2001inferring}, therefore, AS-GCN has the ability to highly summarize the training set with relevant feature information and generalize it to the test set. In particular, the main contributions of this paper are summarized as follows:

\begin{enumerate}[(1)]
    \item  We focus on the multi-classification of the multiple relationships between ASes for the first time. Meanwhile, we aggregate multiple superior algorithms based on the idea of hard voting to obtain the more comprehensive dataset, which well solves the deviation and limitation of the training data of the previous research just from a single platform. 
    \item We introduce two new features, \emph{common neighbor ratio} and \emph{AS type}, and develop a new GCN framework AS-GCN to solve this multiple relationships inference problem under complex scene, which takes into account both the global and local topological properties simultaneously. To the best of our knowledge, this is the first trial that use GCN to solve this problem.
    \item Comprehensive experiments show the outstanding performance of our AS-GCN, by comparing with a series of baselines, especially on the more challenging multi-classification task. Such results indicate that AS-GCN could be a better choice in predicting diverse relationships between ASes. 
\end{enumerate}

The rest of paper is organized as follows. Section 2 introduces related work, and describes in detail the inference algorithms proposed in recent years. Section 3 details the source and processing of the data. The challenges of the current research and the introduction of the basic ideas combined with the GCN model are described in Section 4. Section 5 details the design and implementation of our framework AS-GCN. In Section 6, we conduct extensive experiments and hyper-parameter analysis. Section 7 concludes this work.

\section{Related Work}
In this part, we will briefly review the background and the related works on AS relationship and the inference techniques.

\subsection{AS Relationship}
As officially described, an Autonomous System (AS) is commonly defined as a collection of IP prefixes under the control of one or more network operators that presents a common, clearly defined routing policy to the Internet. And the standard inter-domain routing protocol is the Border Gateway Protocol (BGP). The BGP route information is the main data source used to map the AS-level Internet topology, and Route Views~\cite{routeviews} and RIPE Network Coordination Centre (RIPE NCC)~\cite{riperis} collect BGP route information through a set of route collectors (BGP monitors or Vantage Points (VPs)~\cite{orsini2016bgpstream}).

The Internet topology at the AS-level is typically modeled using a simple graph where each node is an AS and each link represents a business relationship between two ASes. These relationships reflect who pays whom when traffic is exchanged between the ASes, and are the key to the normal operation of the Internet ecosystem. Traditionally, these relationships are categorized into (1) Customer-Provider (C2P), (2) Peer-Peer (P2P), and (3) Sibling relationships (S2S)~\cite{motamedi2014survey}, but other forms of relationships are known to exist as well. In a C2P relationship, the customer is billed for using the provider to reach the rest of the Internet. The other two types of relationships are in general settlement-free. In other words, no money is exchanged between the two parties involved in a P2P or S2S relationship.

Understanding of AS relationship is vital to the technical research and economic exchanges of the inter-domain structure of the Internet. The relationship between ASes is regarded as private information by various organizations, institutions, and operators and is not published on the open platform. By considering the Internet as a complex network, various excellent AS relationship inference algorithms have been proposed to predict the AS-level structural relationship of the Internet, which is of particular significance for Internet security. 

\subsection{Inference Techniques}
Gao~\cite{gao2001inferring} first proposed to enhance the representation of the AS graph by defining multiple business relationships, and put forward an assumption, that valid BGP paths are \emph{valley-free}~\cite{qiu2007toward}, (i.e. $[C 2 P / S 2 S]^{n}[P 2 P]^{(0,1)}[P 2 C / S 2 S]^{m}, n \geq 0, m \geq 0$, a path consists of zero or more C2P or S2S links, followed by zero or one P2P links, followed by zero or more P2C or S2S links, the shape is composed of an uphill path and a downhill path or one of the two), which plays an important role in the later process of inference algorithm research. Since then, a series of methods~\cite{di2003computing, dimitropoulos2007relationships, gregori2011bgp} have been proposed to enhance the inference performance by improving \emph{valley-free} feature. However, the subsequent researches proved that only considering \emph{degree} and \emph{valley-free} may be not enough to infer the complex relationships of the Internet. 

Unlike previous approaches, AS-Rank~\cite{luckie2013relationships} does not seek to maximize the number of \emph{valley-free} paths but rely on three assumptions about the Internet inter-domain structure: (1) An AS enters into a provider relationship to become globally reachable; (2) There exists a peering clique of ASes at the top of hierarchy; (3) There is no cycle of P2C links. Based on these assumptions, the features of clique, transit degree and BGP path triplets that meet the practical significance are proposed to realize the inference. Due to its high accuracy and stability, AS-Rank has been used on CAIDA~\cite{caida} until now. 

Giotsas et al.~\cite{giotsas2014inferring} rethought the problem of inferring complex AS relationships proposed in \cite{luckie2013relationships} and presented a new algorithm to infer the two most common types of AS relationships: \emph{hybrid relationships} and \emph{partial transit relationships}.  Extending the use of BGP, traceroute, and geolocation data, their inferences achieved 92.9\%  and 97.0\% positive predictive values on hybrid and partial transit relationships, respectively. 

ProbLink~\cite{jin2019stable} is the first probabilistic AS relationship inference algorithm based on Naive Bayes, to reduce the error rate and overcome the challenge in inferring hard links, such as non-valley-free routing, limited visibility, and non-conventional peering practices. This approach demonstrates its practical significance in detecting route leak, inferring complex relationships and predicting the impact of selective advertisements. TopoScope\cite{jin2020toposcope} further uses ensemble learning and Bayesian Network to reduce the observation bias, and reconstruct Internet topology by discovering hidden links.

Varghese et al.~\cite{susan2015machine} uses AdaBoost ~\cite{friedman2000additive} algorithm to train a model that predicts the link types in a given AS graph using two node attributes - degree and minimum distance to a \emph{Tier-1} node, but neither the choice of data set nor the setting of the experimental group are rigorous enough. Shapira et al.~\cite{shapira2020unveiling} refers to Natural Language Processing (NLP) to propose a deep learning model \emph{BGP2VEC}. In recent years, many methods have mainly focused on the inference of the relationship P2P and P2C. Research by Giotsas et al.~\cite{giotsas2014inferring} showed the interconnections of ASes of a real Internet is much more complex and diverse. Therefore, it is necessary to propose more suitable algorithms for inferring complex relationships, so as to better understand the increasingly large Internet topology.

\section{Data Set}
In this section, we elaborate on the source of the experimental BGP paths data, and introduce the standard dataset for labeled S2S and X2X relationship. On the validation dataset, we select the same data set as many algorithms by focusing on community attribute\cite{giotsas2012valley}, and our voting-based training dataset is more representative.
\subsection{BGP Paths}
We collect BGP paths from Route Views~\cite{routeviews} and RIPE NCC~\cite{riperis}, the most popular projects managing route collectors and making their dumps accessible and usable by any researcher. Currently, these two projects manage 29 and 25 collectors, respectively, with more than 1000 \emph{vantage points} (VPs) in total (this number is growing over time). For each VP, its collector dumps a snapshot of the Adj-RIB-out table at the frequency of few hours (\emph{RIB dump}) and few minutes (\emph{Updates dump}) ~\cite{orsini2016bgpstream}, respectively. In order to make the subsequent experimental results universal and the evaluation credible, we download RIB files from the first day of April, August, and December every year from 2012 to 2020 and extract the BGP paths that announce reachability to IPv4 prefixes.

Because the dataset comes from the real-time acquisition of the Internet, some necessary preprocessing needs to be done before the experiment. We first use the table provided by Internet Assigned Numbers Authority (IANA)~\cite{iana} to remove the paths that have numbers that are not allocated to Regional Internet Registries (RIRs)~\cite{rirs}. We also sanitize the BGP path~\cite{katz2011machiavellian} containing AS loops, when the same AS number (ASN) in a path is not adjacent. And we compress the paths that have the same consecutive ASN (i.e. from "A B C C" to "A B C").

\subsection{S2S Relationship}
We use CAIDA's AS-to-Organization mapping dataset~\cite{as2org}, which applies WHOIS information available from Regional and National Internet Registries. Mapping is available from October 2009 onwards and the new mapping is added in each quarter. Our frequency of obtaining BGP path data is similar to the updating cycle. This dataset contains the ASN and the organization it belong to, so we could infer the links that its endpoints managed by the same organization as sibling relationship (S2S relationship). In the following binary classification experiment, we will preprocess the dataset as known, and in the multi-classification experiment, we will randomly split it into train, validation and test set.

\begin{table*}[]
\caption{Training dataset established from the intersection of the three methods.}
\renewcommand\arraystretch{1.3}     
\centering
\renewcommand{\multirowsetup}{\centering}
\begin{tabular}{c|c|c|c|c|c|c|c}
\bottomrule[2pt]
\multirow{3}{1.5cm}{Date} & \multirow{3}*{Total links} & \multicolumn{3}{c|}{Real Situation} & \multirow{3}*{Intersection} & \multirow{3}*{Coincidence rate} & \multirow{3}*{Intersection Acc}\\
\cline{3-5}
&&\multirow{2}{1.8cm}{AS-Rank}  &\multirow{2}{1.8cm}{ProbLink}  &\multirow{2}{1.8cm}{TopoScope}&&
\\&&&&&&\\
\hline
2016 & 265633 & 254297 & 254466 & 258538 & 221272 & 83.30\% & 98.13\%\\
\cline{1-8}
2017 & 255694 & 246117 & 249752 & 249755 & 214465 & 83.88\% & 97.06\%\\
\hline
2018 & 263219 & 254554 & 254700 & 255352 & 214281 & 81.97\% & 99.26\%\\
\toprule[2pt]
\end{tabular}
\label{Tab1}
\end{table*}

\subsection{IXP List}
An Internet Exchange Point (IXP) is a physical infrastructure used by Internet Service Providers (ISPs) and Content Delivery Networks (CDNs) to exchange Internet traffic between their ASes. An IXP can be distributed and located in numerous data centers (also known as facilities), and a single facility can contain multiple IXPs. An AS connected to a given IXP is known as a member of that IXP. Our IXP list is derived from visiting PeeringDB~\cite{peeringdb} for networks of type ``\emph{Route Server}'' and extracting the ASN. Since IXP provides a neutral shared exchange structure, and clients can exchange traffic with each other after establishing a peer-peer connection, previous works remove BGP paths contained IXPs to study relationship between unneutral ASes instead of relationship between IXPs and unneutral ASes. In this paper, the connection between IXP and other ASes is defined as X2X relationship, and we expect to recognize this kind of relationship through some empirical analysis. Although not all IXPs have route servers, the number of AS contained in the newly extracted list can be considered as the lower bound. There were 336 IXP ASes in this list on 12/01/2020.

\subsection{Experimental dataset}
\textbf{Training Dataset:} We obtain BGP paths from Route Views and RIPE NCC on the first day of April in 2014-2018 as our source data. After unified preprocessing, three series of link relationship inference data containing labels are obtained using three existing inference techniques that have been verified by a large number of experiments. Using the intersection of the result sets obtained by the three inference methods on the same day as the experimental data for subsequent experiments. The reason is as follows:
\begin{enumerate}[ (1) ]
    \item  The accuracy of each relationship inference approach is generally more than \emph{90\%} in recent years, which represents that the existing methods have mastered the general features of the Internet topology structure. Similar to previous study~\cite{shapira2020unveiling}, only using the inference result of a certain method as a data set must have a bias which cannot be ignored.
    \item The scale of the Internet is growing, and the number of route collector and VP is also increasing. This means that the data in recent years is better than before in terms of quantity and structural integrity. The amount of data determines the model's ability to express the real situation to a certain extent.
    \item The intersection of the inference results of multiple technologies can be seen as a method of hard voting based on probability and our vote focuses on the result consistently. Based on the results of previous researches, we have greatly reduced the error of the dataset we constructed.
\end{enumerate}

TABLE~\ref{Tab1} shows the results of BGP paths in the same time period under the current three advanced inference models, as well as their number of intersections and proportions.

\textbf{Validation Dataset:} The community attribute~\cite{giotsas2012valley} is an optional BGP route tag that simplifies the implementation of routing policy. This attribute is widely used by ASes to encode metadata on a route, which often includes relationship information. Although the usage of BGP community is not standardized, the IRR~\cite{irrs} provides the document to decode the meanings of community values, and make it possible to infer either the type of AS relationship with the neighbors. Our dataset collected by ProbLink~\cite{jin2019stable}, which contains same labeled datasets based on community attribute from 2014 to 2018 in April each year. Same as the previous series of work~\cite{luckie2013relationships, jin2019stable, jin2020toposcope}, we treat this validation set as "\emph{best-effort}" dataset to reproduce the existing inference models and evaluate our AS-GCN model.

\begin{table}[h!]
\caption{Validation dataset.}
\renewcommand\arraystretch{1.3}
\begin{tabular}{ p{1.5cm}<{\centering}|p{2.8cm}<{\centering}|p{1.3cm}<{\centering}|p{1.5cm}<{\centering} } 
\bottomrule[2pt]
Date & Validation (P2C/P2P) & All links & (S2S/X2X)\\
  \hline
  04/01/2014 & 44474(26210/18264) & 146167 & 1663/1164 \\ 
  \hline
  04/01/2015 & 42512(26528/15984) & 169019 & 1906/1881 \\ 
  \hline
  04/01/2016 & 49074(28214/20860) & 221573 & 2133/3957 \\ 
  \hline
  04/01/2017 & 49604(29362/20242) & 219713 & 2276/2686 \\ 
  \hline
  04/01/2018 & 44868(30284/14584) & 223221 & 2760/4467 \\ 
\toprule[2pt]
\end{tabular}
\end{table}

Table 2 shows the details of this validation dataset from 2014 to 2018. 
The number of real routers (VPs) that can establish BGP peering sessions~\cite{orsini2016bgpstream} is increasing, allows us to obtain more BGP paths. Simultaneously, the more complete community feature description document provided by PeeringDB~\cite{peeringdb} also expands the extraction of the validation set. We can see that the total number of AS links, the numbers of AS links in the validation sets and diverse relationship links (i.e., S2S and X2X) are showing an increasing trend.

\section{Challenges and Basic Ideas\label{Sec4}} 
In this section, we discuss the major challenge of current inference algorithms, and then put forward our basic ideas. In particular, we conducted a quantitative analysis using 04/01/2017 dataset.

\begin{figure}[hpt]
\centering
{
    \begin{minipage}[b]{\linewidth}
        \centering
        \includegraphics[scale=0.35]{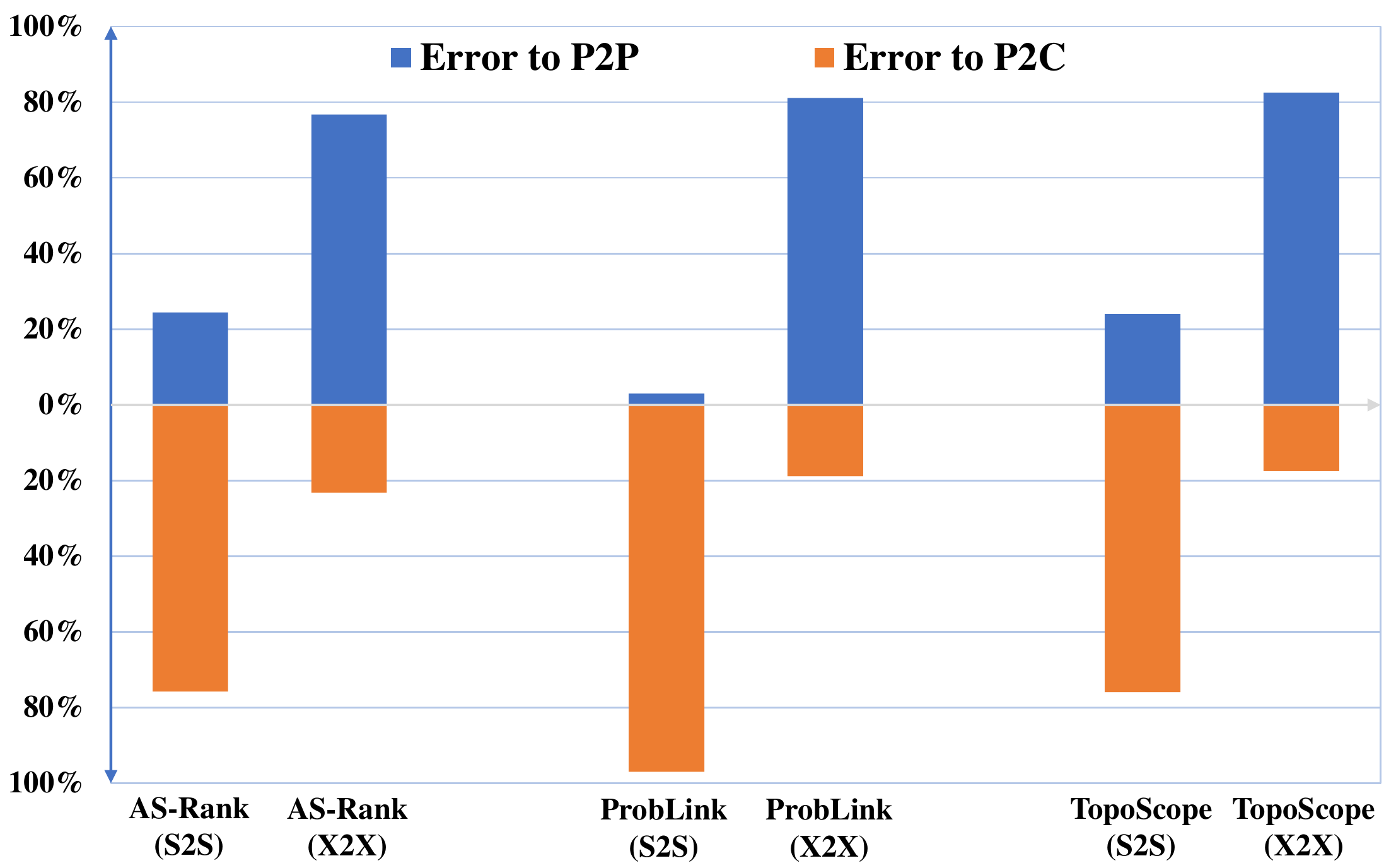}
    \end{minipage}
}
\caption{The fraction of misclassification of complex relationships by different algorithms.}
\label{Fig1}
\end{figure}

\subsection{AS Relationship Classification}
Traditionally, Internet topological graph is constructed based on the corresponding relationship between different organizations. The most frequently considered business relationships that need to be inferred are P2P (cooperation and mutual benefit) and P2C (service is accompanied by money exchanges). While S2S (mentioned in section 3.2 above) and X2X (connect to IXP for more cost-effective communication) relationships are often considered to be known apriori.

The S2S relationship is not highly distinguishable from other types, and we conduct experiments to evaluate the inference of the existent algorithms for special types. Fig.~\ref{Fig1} gives a histogram of the misclassification of the two kinds of relationships S2S and X2X by different algorithms on 04/01/2017 dataset. We can clearly see that the trends of the three mainstream algorithms are consistent with each other: more than 75\% of S2S links will be classified as P2C, while more than 75\% of the X2X links will be classified as P2P. For instance, 2387 S2S links are inferred by the ProbLink method, 2316 of which (nearly 97\%) were classified as P2C. Such results indicate the strong similarity between S2S and P2C, and also between X2X and P2P, making it quite difficult to distinguish them by using the current methods.

We argue that the characteristics of communication mode within the organization and the role of the IXP are all necessary for a better understanding and fine management of the Internet. According to what we have observed, it is difficult to only consider a several appropriate features to distinguish four types of AS relationship for multi-classification problems. Thus, we need to design an automatic learning algorithm to solve this problem.

\subsection{Aggregate Surrounding Information}
Triplet feature~\cite{luckie2013relationships} is used to simulate the \emph{valley-free} path in a probabilistic way. Although it has been proven that there is a non-negligible part of the real BGP paths that violates the \emph{valley-free} path~\cite{giotsas2012valley}, \emph{valley-free} feature has played an important role in many methods since it was proposed in 2001~\cite{gao2001inferring}. The advantage of using triplet instead of focusing on the entire path is allowing us to ignore non-hierarchical segments of paths and pay more attention to the neighboring links. Typically, a BGP path can be decomposed into adjacent link pairs or three consecutive links. For example, “$6939|4826|38803|56203$” can be decomposed into link pairs: “6939 4826 38803”, “4826 38803 56203”, and decomposed into three consecutive links: “$Null$-$\langle6939, 4826\rangle$-$\langle4826, 38803\rangle$", "$\langle6939, 4826\rangle$-$\langle4826, 38803\rangle$-$\langle38803, 56203\rangle$", \quad"$\langle4826, 38803\rangle$-$\langle38803, 56203\rangle$-$Null$”.

\begin{figure}[hp]
\centering
{
    \begin{minipage}[b]{\linewidth}
        \centering
        \includegraphics[scale=0.35]{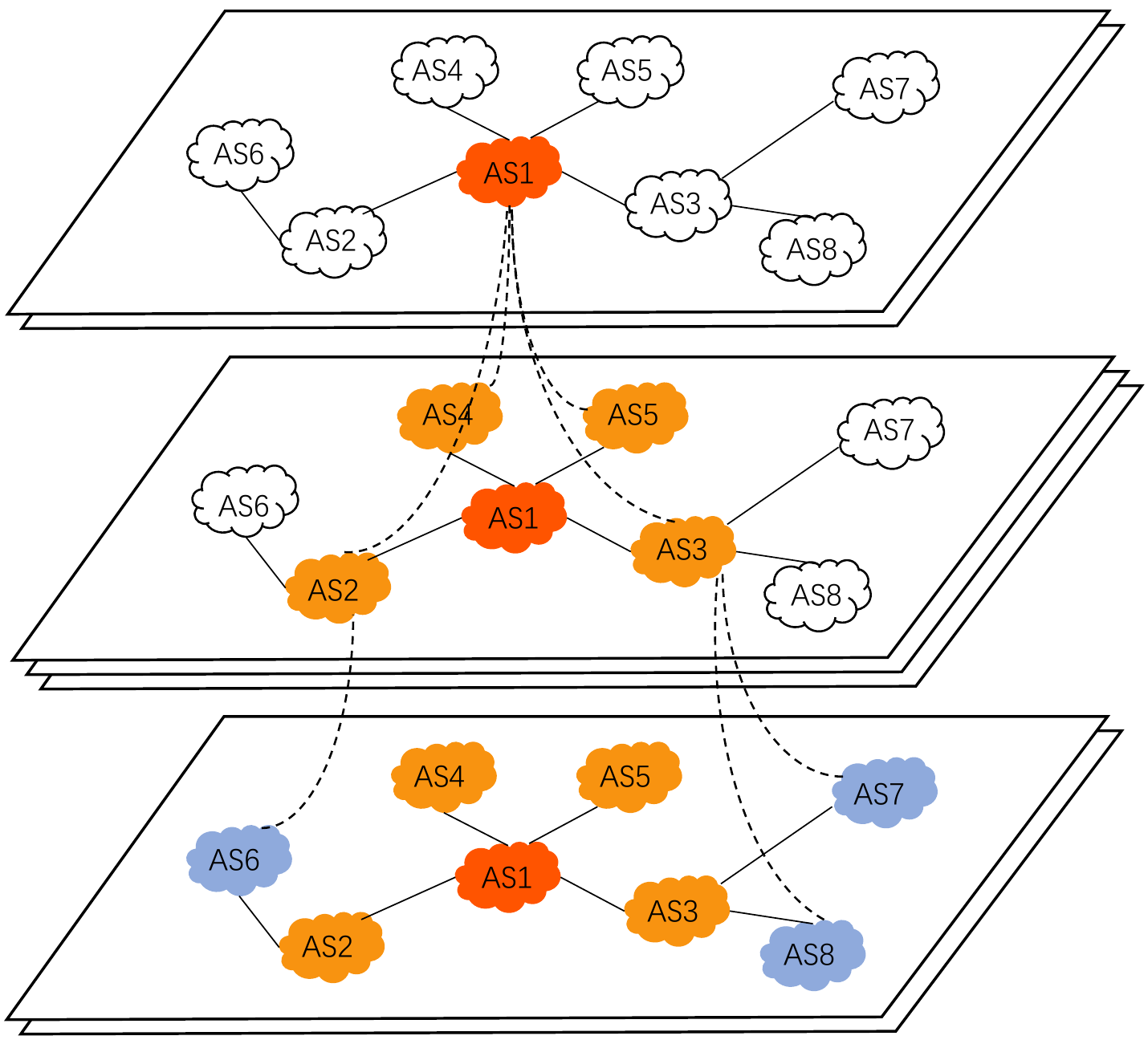}
    \end{minipage}
}
\caption{Schematic diagram of the aggregating surrounding information process of the GCN model.}
\label{Fig2}
\end{figure}

Naturally, the triplet features are local properties based on the first-order neighbors of ASes in the Internet graph. In this sense, we could benefit by adopting Graph Convolutional Network (GCN)~\cite{kipf2016semi} to utilize more structural information beyond first-order neighbors for classification. 
From the row vector perspective of matrix multiplication, graph-based convolution process is equivalent to the aggregation operation of the feature vectors of the neighbor nodes. 

The operation of such aggregating neighbor information is given by the schematic diagram, as shown in Fig.~\ref{Fig2}. The process of graph convolution operation can realize efficient filtering operation on graph data, and GCN brings powerful fitting capabilities by stacking and modifying multiple GCN layers. The model structure will be elaborated in the algorithm design in the next section.

\section{Methodology} 
In this section, we formally propose AS-GCN, a graph convolutional network based framework, to infer the AS relationship under complex Internet structures automatically and adaptively. The framework runs in two steps. First, we extract eight features of the nodes (ASes) and the edges (links) which are highly distinguishable and relevant from the perspective of graph structure. Second, we input graph structure and feature information into our model for training and comparing with ground truth to minimize the negative log-likelihood loss, so as to solve binary and multi-class AS relationship classification problems. 

\subsection{Features Design and Analysis}
Before feature analysis, we calculate the difference of features between the target AS and the source AS to represent the corresponding link:
\begin{equation}
\Delta =  | f_{AS1} - f_{AS2} |
\label{Eq2}
\end{equation}
where $f_{AS1}$ and $f_{AS2}$ represent the feature values of AS1 and AS2, respectively, and $\Delta$ is the difference. This method is used by default in the feature analysis below. 

\subsubsection{Degree and Transit Degree}
\emph{Degree} is one of the most basic indices in graph theory to describe the importance of a node. The degree of node $i$ in an undirected network is defined as the number of edges directly connected to it. \emph{Transit degree}~\cite{luckie2013relationships} is the number of unique neighbors appearing in transit paths of an AS by extracting triplets. For example, suppose existing a path \emph{AS1, AS2, AS3, AS4}, there are two triplets \emph{(AS1, AS2, AS3), (AS2, AS3, AS4)}. Both \emph{AS2} and \emph{AS3} have two different neighbor nodes from the extracted triples, so their \emph{transit degrees} are both 2. Besides, ASes with a transit degree of zero are \emph{stub} ASes, which are at the outermost layer of the network (i.e. \emph{AS1}, \emph{AS4}). The \emph{transit degree} is more suitable for describing the relationship clusters of ASes in the Internet, which mainly reflects the scale of the customer service and cooperation of an AS. Therefore, \emph{degree} and \emph{transit degree} are used as important graph structure features for subsequent algorithm design. 

\begin{figure*}[ht]
\centering
{
    \begin{minipage}[b]{\linewidth}
        \centering
        \includegraphics[scale=0.9]{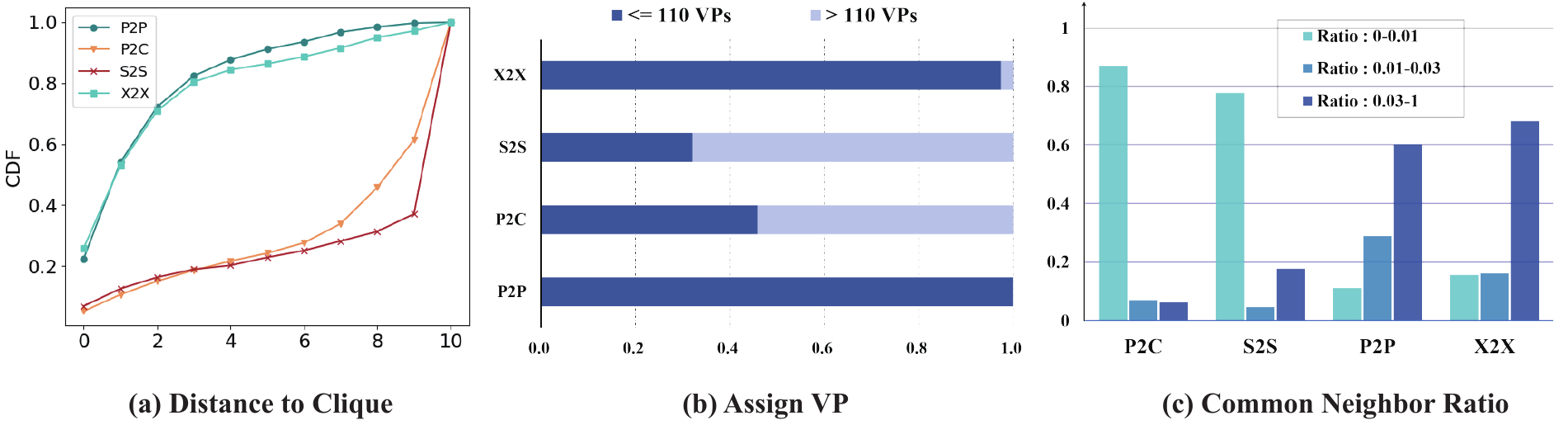}
    \end{minipage}
}
\caption{Analysis of the \emph{Distance to Clique}, \emph{Assign VP}, \emph{Common Neighbor Ratio}. (a) CDF of absolute distance between ASes to clique for different relationships. (b) The distribution of the number of VPs with a threshold 110 that can be detected on each relationship's links. (c) The distribution of different common neighbor rates for different relationships.}
\label{Fig3}
\end{figure*}

\begin{figure*}[ht]
\centering
{
    \begin{minipage}[b]{\linewidth}
        \centering
        \includegraphics[scale=0.85]{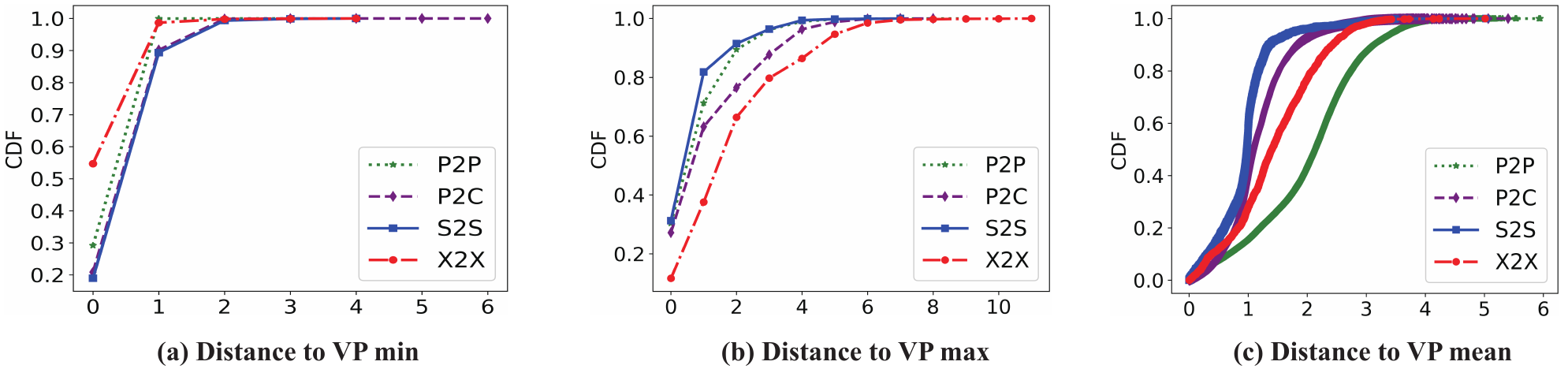}
    \end{minipage}
}
\caption{CDF of the \emph{Distance to VP} using different calculation methods on four types of AS relationship.}
\label{dis2vp}
\end{figure*}

\subsubsection{Distance to Clique}
Inferring clique, the ASes exist at the top of the hierarchy and form a transit-free state with each other (can be seen as P2P relationship), is the first step of extracting the \emph{distance to clique} feature. About the detailed steps of selecting clique ASes, we refer to the previous work~\cite{luckie2013relationships}. The \emph{distance to clique} feature is mainly to capture the distance from the ASes to the network center. This feature is based on such an assumption: High-tier ASes generally has a large number of customers, so they are easier to be peers to achieve mutual benefit, and the better strategy for the ASes at the low-tier are to rely on the top ASes to achieve global accessibility and form provider-customer (P2C relationship) finally.

We first construct an undirected graph by extracting the AS links in the BGP paths as edges. After that, we calculate the average shortest path distance from each AS to the clique ($\frac{\sum_{i=1}^{N} D_{i}}{N}$, $N$ is the number of the clique, and $D_{i}$ is the shortest distance from an AS to the ${i}^{th}$ AS in the clique) and map it to the corresponding link using Eq. (\ref{Eq2}). Finally, we use the 04/01/2017 validation dataset to evaluate the four types of AS links, and the distribution of \emph{distance to clique} is shown in Fig.~\ref{Fig3} (a). This feature well reflects the significant difference between P2P and P2C relationship, as well as the similarity of the two pairs of AS relationships (i.e., P2C and S2S, P2P and X2X). This is one of the reasons that the X2X and S2S relationships are difficult to distinguish using previous inference algorithms, which has been described in Section 4.1.

\subsubsection{Assign VP}
Vantage Points (VPs, VP can be intuitively understood as the first node of the AS path.) are typically distributed in many different geographic locations, especially at the upper tiers of the Internet hierarchy. Meanwhile, the number of VP is also very limited, compared with the scale of complete Internet structure. Here, we analyze the quantity of VPs, which can detect the same AS link. We visualize the discrimination among different types of AS relationship in Fig.~\ref{Fig3} (b) about this feature. From Fig.~\ref{Fig3} (b), we can observe that more than 97\% P2P links and X2X links can be detected by less than 110 VPs (refer to the previous work~\cite{jin2020toposcope}), while more than half P2C and S2S links are seen by more than 110 VPs. Hence, for the single feature \emph{Assign VP}, the two pairs of AS relationships (i.e., P2C and S2S, P2P and X2X) tend to be similar. This result once again confirms the challenge of multi-class relationship classification problem.

\subsubsection{Common Neighbor Ratio}
This feature is defined as the ratio of common neighbors between ASes to the total number of neighbors, which is a new key point that we combined with the practical significance of the graph structure. Our intuitive idea is as follows: due to business, ASes belonging to the same organization (sibling relationship) are unlikely to connect to other types of ASes at the same time. In addition, IXPs are used to promote the interconnection and exchange of the backbone of the Internet and connect many important ASes because of its functions. Therefore, IXPs' common neighbor ratio will be higher than other types of ASes. As shown in Fig.~\ref{Fig3} (c), the proportion of P2P and X2X relationship with large common neighbor ratios both exceed 60\% which validates our intuitive idea. However, the two pairs of AS relationships (i.e., P2C and S2S, P2P and X2X) are also indistinguishable up to \emph{common neighbor ratio} which well reflects the challenge of the multi-class relationship classification problem proposed by us for the first time.

\subsubsection{Distance to VP}
Different from \emph{distance to clique}, we also pay attention to the distance from each node (AS) to the first node (called Vantage Point, VP) in each BGP path. This feature indicates that we expect to count the distance set from the target AS to VP in all BGP paths to reflect the position of a link in all paths (i.e., front, middle or end) and it has a certain correlation with the AS relationship. Because the same node will appear in several paths, the \emph{distance to VP} value of the node will be expressed as a set of integers. In the face of these integer sets, the mean value of the set represents the universality of the node position, and the maximum and minimum values of the set represent the specificity of the node position. As shown in Fig.~\ref{dis2vp}, we can observe that using the average value of \emph{distance to VP} is more discriminative and dense among the four types. In the following feature importance analysis (see TABLE~\ref{Tab5}), it can also be fully proven that the importance of the mean value is higher than the maximum and minimum values.

\subsubsection{Node Hierarchy}
The Internet obeys a certain hierarchical structure~\cite{carmi2007model}. Considering the hierarchical features of ASes, we pay attention to the distribution of different types of AS links in the Internet topology. We refer to the work of \emph{K-Shell}~\cite{carmi2007model}, using \emph{transit degree} to decompose the Internet into three components (as shown in Fig.~\ref{Fig4}):

\begin{figure}[ht]
\centering
{
    \begin{minipage}[b]{\linewidth}
        \centering
        \includegraphics[scale=0.4]{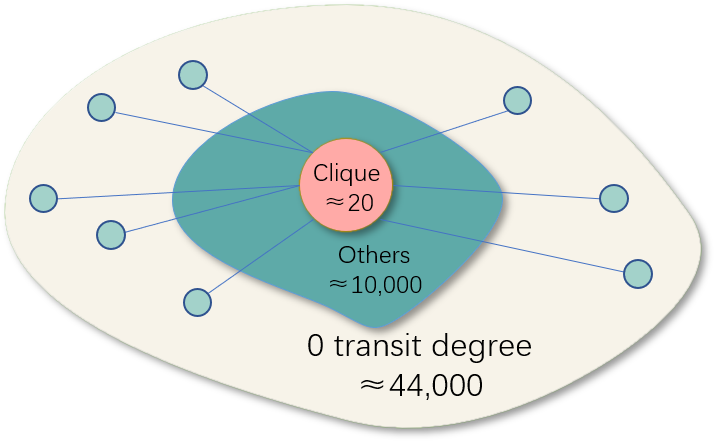}
    \end{minipage}
}
\caption{The hierarchical structure of the AS-level Internet topology.}
\label{Fig4}
\end{figure}

\begin{enumerate}[(1)]
    \item All ASes in the clique form a nucleus, which belong to the smallest clusters. However, the ASes in the clique have a large average degree, and most of AS links which started from the clique connect to the outer structure of Internet topology.
    \item The rest of the ASes with zero transit degree is considered as shell of the Internet. Simultaneously, there are cases where some ASes are directly connected to the core. This part constitutes the largest component.
    \item The remaining ASes are all classified into one category.
\end{enumerate}

\subsubsection{AS Type}
\begin{figure}[ht]
\centering
{
    \begin{minipage}[b]{\linewidth}
        \centering
        \includegraphics[scale=0.4]{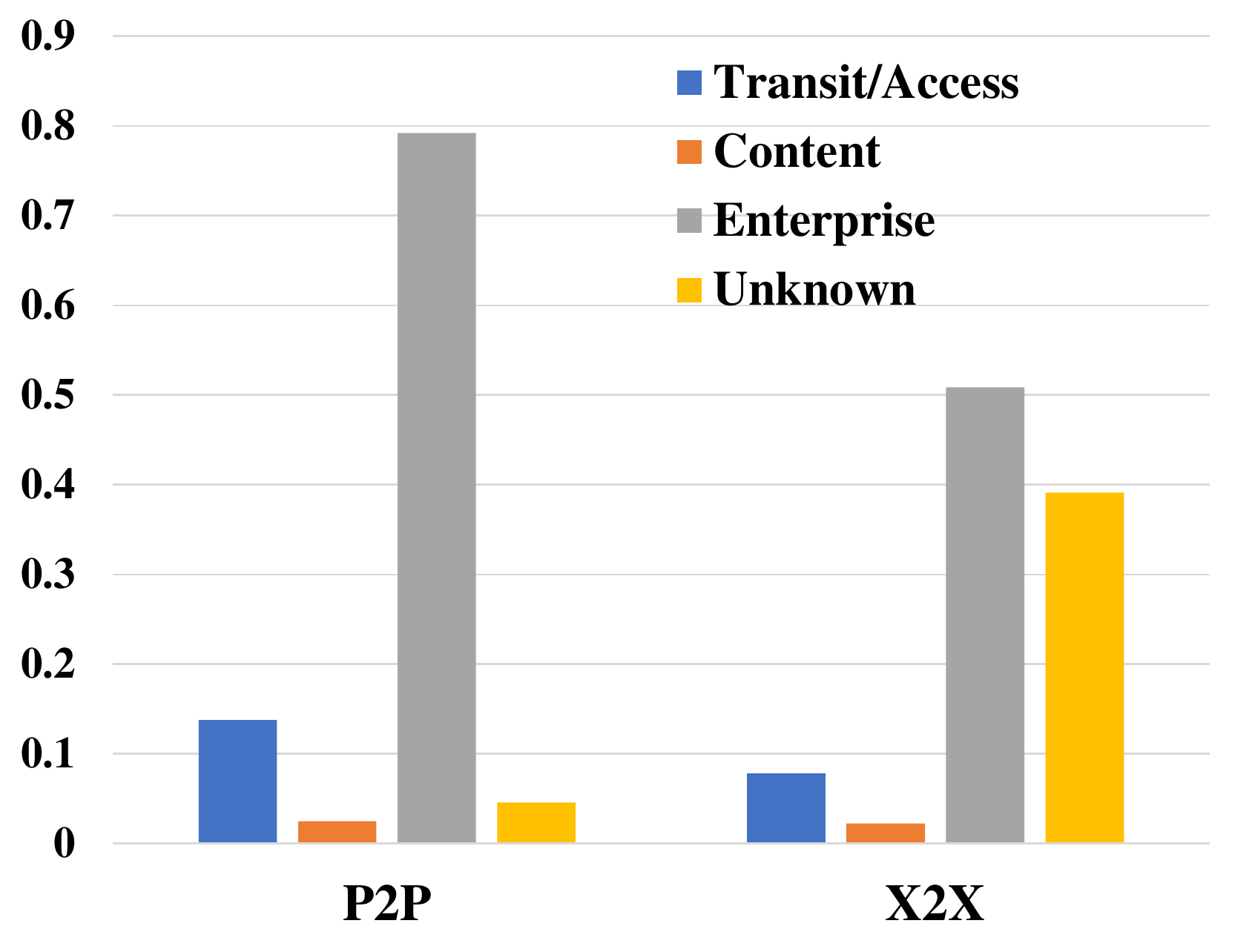}
    \end{minipage}
}
\caption{The distribution of the four AS types between P2P and X2X.}
\label{Fig5}
\end{figure}

AS as an organization has related business types due to its functions. \emph{AS type} has been considered to be a very important feature, because from the practical significance, it has a direct impact on the AS relationship. We get the AS classification dataset from CAIDA~\cite{as_classification}. The ground-truth data is extracted from the self-reported business type of each AS list in PeeringDB~\cite{peeringdb}. After that, \emph{AS type} can be summarized into three main categories: (1) \emph{Transit/Access}. This type of ASes are inferred to be either a transit and/or access provider. (2) \emph{Content}. This type of ASes provide content hosting and distribution systems. (3) \emph{Enterprise}. This type of ASes include various organizations, universities and companies. Furthermore, we also add the fourth type: (4) \emph{Unknown}. This group contain those ASes that don't have a clear type, and the neutral ASes (i.e., IXP) that does not belong to the first three categories. In the experiment, it is found that 60\% of X2X edges contain ASes of the \emph{Unknown} type, which is well distinguished from the other three types of ASes as shown in Fig.~\ref{Fig5}.

\subsection{Model Framework}
 AS-GCN mainly conists of five types of layers, i.e., Input layer, Feature layer, GCN layer, MLP layer, and Output layer.

\begin{figure*}[ht]
\centering
{
    \begin{minipage}[b]{\linewidth}
        \centering
        \includegraphics[scale=0.39]{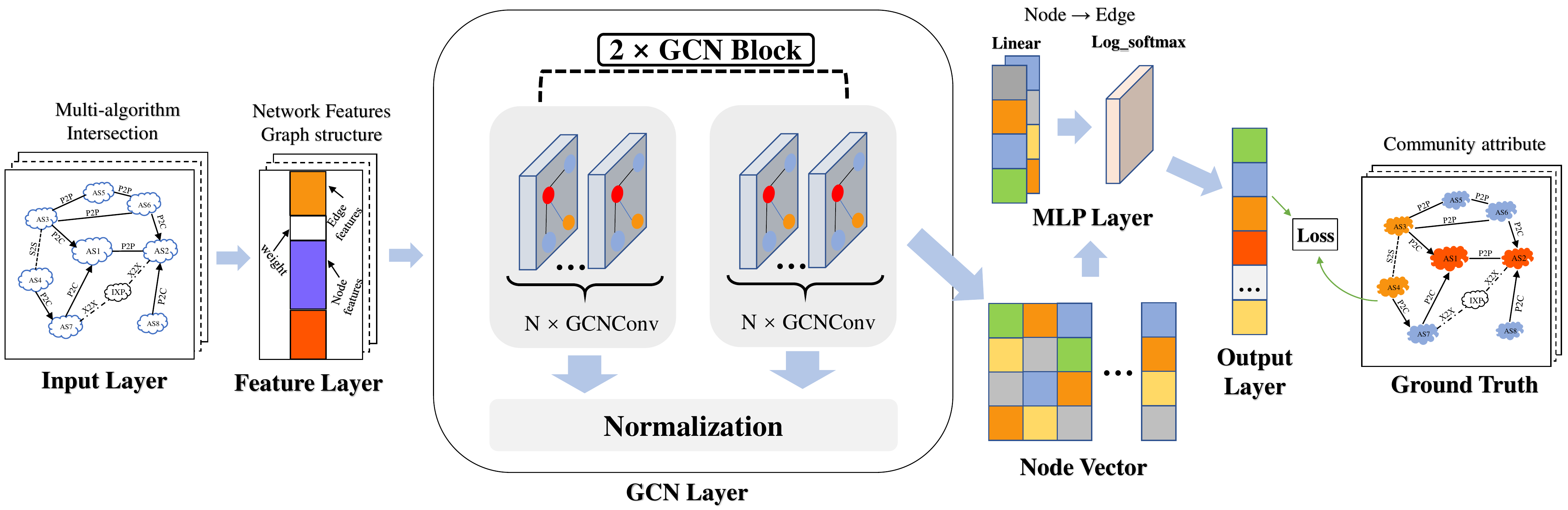}
    \end{minipage}
}
\vspace*{-4mm} 
\caption{Model Framework of AS-GCN.}
\label{Fig6}
\end{figure*}

\textbf{Input Layer:} The source dataset of the input has elaborated in section 3.4. Meanwhile, the links of the S2S and X2X relationship in the dataset are also calibrated through the public information provided by CAIDA. For binary classification problem (i.e., P2P and P2C), the other two types of links (i.e., S2S and X2X) would be removed from the dataset in advance. Given an undirected and unweighted graph $G=(V, E)$, which represents a static Internet AS-level network, where $V$ denotes the set of ASes and $E\in{V\times{V}}$ denotes the set of business relationships between ASes. Let $v_i\in{V}$ denotes an AS and $e_{ij}=(v_i, v_j)\in{E}$ denotes an AS link pointing from $v_i$ to $v_j$. The adjacency matrix $\textbf{A}$ is a $n\times{n}$ matrix with $\textbf{A}_{ij}=1$ if $e_{ij}\in{E}$ and $\textbf{A}_{ij}=0$ if $e_{ij} \notin {E}$. The graph has node attributes $\textbf{X}$, where $\textbf{X} \in {\mathbb{R}^{n\times{d}}}$ denotes the feature matrix and $x_v\in{\mathbb{R}^d}$ represents the feature vector of node $v$.

\textbf{Feature Layer:} As illustrated in Fig.~\ref{Fig6}, the feature layer is mainly to construct its own feature vector for each node and generate a corresponding weight value for each edge. Of course, the feature vector is composed of the seven important features mentioned above (i.e., \emph{degree}, \emph{transit degree}, \emph{distance to clique}, \emph{distance to VP}, \emph{assign VP}, \emph{node hierarchy}, and \emph{AS type}) and the related features after normalization. \emph{common neighbor ratio} constitutes the weight of the edges.

\textbf{GCN Layer:} In the previous section, we have introduced the basic idea of using graph convolution operation. GCN is a semi-supervised learning algorithm for graph structured data, which makes use of Laplace transform to make the node aggregate the features of higher-order neighbors. In particular, GCN instantiates $\textbf{A}$ to be a static matrix closely related to the normalized graph Laplaican to get the following expression:
\begin{equation}
\hat{\textbf{A}}=\tilde{\textbf{D}}^{-1 / 2} \tilde{\textbf{A}} \tilde{\textbf{D}}^{-1 / 2}
\end{equation} 
where $\tilde{\textbf{A}}=\textbf{A}+\textbf{I}$, $\tilde{\textbf{D}}_{ii}=\sum_j\tilde{\textbf{A}}_{ij}$, with $\textbf{A}$ being the adjacency matrix and $\textbf{I}$ being the identity matrix. $\hat{\textbf{A}}$ can be regarded as a graph displacement operator.

GCN carries out convolution operation in the spectral domain, and each operation can aggregate an additional layer of features. Spectral convolution function is formulated as

\begin{equation}
\textbf{H}^{(l+1)}=\sigma\left(\hat{\textbf{A}} \tilde{\textbf{H}}^{(l)} \textbf{W}^{(l)}\right)
\end{equation}

\noindent where $\textbf{W}^{(l)}$ denotes the layer-specific trainable weight matrix and $\sigma\left(\cdot\right)$ is a non-linear activation function (i.e., Sigmoid function). Moreover, $\textbf{H}^{(l)}\in{\mathbb{R}^{n\times{k}}}$ represents the matrix of activation at $l$ layer, where $n$ and $k$ denote the number of nodes and output dimensions of layer $l$, respectively. In particular, we set $\textbf{H}^{(0)}=\textbf{X}$ as initialization input. Each convolution operation would capture the neighbor’s features of additional layer. If the objects of the first matrix multiplication are $\textbf{A}$ and $\textbf{X}$, then they are equivalent to the nodes combined first-order neighbor features. The more such multiplications, the more layers of information that are abstractly merged.
 
In the following, we design a model, which contains two identical block, each with two-layer GCN, to achieve semi-supervised node classification on a graph with a symmetric adjacency matrix $\textbf{A}$. Our forward model then takes the block's simple form:
\begin{equation}
f(\textbf{X}, \textbf{A})=\operatorname{Norm}\left(\operatorname{ReLU}\left(\hat{\textbf{A}} \operatorname{ReLU}\left(\hat{\textbf{A}} \textbf{X} \textbf{W}^{(0)}\right) \textbf{W}^{(1)}\right)\right)
\end{equation}
\noindent where $\textbf{W}^{(0)}$ is an input-to-hidden weight matrix and $\textbf{W}^{(1)}$ is a hidden-to-output weight matrix. $\operatorname{ReLU}$ is the activation function and normalize the result. The neural network weights $\textbf{W}^{(0)}$ and $\textbf{W}^{(1)}$ are trained using gradient descent.

\textbf{MLP Layer:} Through the Multilayer Perceptron (MLP) layer, the information aggregated by the node is transformed into the corresponding edge. Let $\textbf{Z}$ denote the output of GCN layer, we can get the following expression: 
\begin{equation}
Y_{ij}=\operatorname{Log\_softmax}\left(\operatorname{Linear}\left(\textbf{Z}_{i}, \textbf{Z}_{j}\right)\right)
\end{equation}
where $Z_{i}$ and $Z_{j}$ denotes the output vector of the two end nodes corresponding to the edge $e_{ij}$ in the graph through the GCN Layer, respectively. $Y_{ij}$ is a probability vector which is used to infer the label of the corresponding edge. Compared with the method of dot product to map edges, MLP generates a vector representation for each edge and is more suitable for classification tasks.

For semi-supervised multi-classification, we apply the cross entropy loss as the objective function, which is formulated as:
\begin{equation}
\mathcal{L}(\theta)=-\sum_{e_{ij} \in E} y_{ij} \log \left(Y_{ij}\right)
\end{equation}

\noindent where $y_{ij}$ denotes the label of edge $e_{ij}$, and $\theta$ denotes all parameters needed to be learned in the model.

\textbf{Output Layer:} The model output is compared with ground truth to minimize the negative log-likelihood loss. In the course of the experiment, the training set is used to train the model to determine the weight and bias of the model, the validation set makes the model in the best state by adjusting the hyperparameters, the test set used only once during the entire experiment is used to evaluate the performance of our model. We save the model with the highest experimental result.

\section{Evaluation}
In this section, we evaluate our GCN-based inference algorithm, AS-GCN, on the experimental dataset. We have not only compared with the stable and accurate inference algorithms proposed before such as AS-Rank, ProbLink and TopoScope, but also compared the performance with machine learning methods only considering proposed features, such as Random Forest, Xgboost, LightGBM. We have proved the superiority of AS-GCN in the following three aspects:
\begin{enumerate}[ (1) ]
    \item Under snapshots at different times, the accuracy of AS-GCN on binary-classification can be maintained above 94\%, which is comparable with many superior algorithms, such as AS-Rank, ProbLink and TopoScope.
    \item The overall accuracy of multi-classification tasks evaluated on equal proportions of data sets can even reach higher than 95\%. Precision and recall also show a general advantage.
    \item  Through the parameter optimization and feature importance analysis of the model, the performance of the model can be explained to a certain extent. 
\end{enumerate}

\subsection{Baseline Methods} 

In order to evaluate the effectiveness of the model, our model is compared with the six different methods falling into two categories: (1) relationship inference methods including AS-Rank, ProbLink and TopoScope; (2) feature-based methods including Random Forest, Xgboost and LightGBM. These baselines are described in the following:
\begin{enumerate}[1)]
\item \textbf{AS-Rank} ~\cite{luckie2013relationships}. The state-of-the-art AS relationship inference technique, called the “AS-Rank” algorithm, it takes 11 intricate steps to label each link as customer-provider (abbreviated as C2P or P2C depending on the directionality of the relationship) or peer-peer (P2P).
\item \textbf{ProbLink} ~\cite{jin2019stable}. ProbLink is the first explainable probability model to infer AS relationships. It uses the general Naive Bayes framework to incorporate multiple link features for inferring.
\item \textbf{TopoScope} ~\cite{jin2020toposcope}. TopoScope is a framework for accurately recovering AS relationships from such fragmentary observations. TopoScope uses ensemble learning and Bayesian Network to mitigate the observation bias originating not only from a single VP, but also from the uneven distribution of available VPs.
\item \textbf{Random Forest} ~\cite{breiman2001random}. Random Forest (RF) contains multiple decision tree classifiers. Through the boot-strap resampling technology, and a random forest composed of multiple classification trees is generated from the continuously training samples. The classification result of the test data is determined by the score formed by the number of votes of the classification tree. 
\item \textbf{Xgboost} ~\cite{chen2016xgboost}. Xgboost is one of the Boosting algorithms. The idea of the Boosting algorithm is to integrate many weak classifiers together to form a strong classifier. Because Xgboost is a boosted tree model, it integrates many tree models to form a strong classifier. The tree model used is the CART regression tree model. 
\item \textbf{LightGBM} ~\cite{ke2017lightgbm}. LightGBM is a new GBDT (Gradient Boosting Decision Tree) implementation with two novel techniques: Gradient-based One-Side Sampling (GOSS) and Exclusive Feature Bundling (EFB). LightGBM speeds up the training process of conventional GBDT by up to over 20 times while achieving almost the same accuracy.
\end{enumerate}

\subsection{Evaluation Metrics}
In order to evaluate the performance of the proposed
model, the following metrics are used: \emph{Recall}, \emph{Precision} and \emph{Accuracy}. The mathematical derivation of these
parameters is illustrated using the below equations:
\begin{equation}
Recall = \frac{TP}{TP+FN}
\end{equation}
\begin{equation}
Precision = \frac{TP}{TP + FP}
\end{equation}
\begin{equation}
Accuracy = \frac{TP+TN}{TP+TN+FP+FN}
\end{equation}

\noindent where \emph{TP}, \emph{TN}, \emph{FP} and \emph{FN} refer to \emph{True Positive}, \emph{True Negative}, \emph{False Positive} and \emph{False Negative}, respectively.

Our model is implemented by PyTorch\cite{paszke2017automatic}. The parameters are updated by Adam algorithm\cite{kingma2014adam}. Each experiment runs 200 epochs in total. In the binary classification, all results are obtained by training the AS-GCN using Adam with weight decay $5 \times 10 ^ {-4}$ and an initial learning rate of 0.1. We use two blocks, where each block has two standard GCN layers (i.e., AS-GCN setting can be represented as $2 \times 2$), to learn the graph structure. In the multi-classification, all results are obtained by training the AS-GCN using Adam with weight decay $0$ and an initial learning rate of 0.05. We use $2 \times 1$ setting of AS-GCN to learn the graph structure. 

In summary, we select the best parameter configuration based on performance on the validation set and evaluate the configuration on the test set.

\subsection{Binary Classification}
As a binary classification task, inferring the relationship between P2P and P2C is the main focus of many existing algorithms. We thus also validate our AS-GCN on the same data set. In particular, our experimental snapshots are taken from the BGP path on the first day of April each year from 2012 to 2018, and our validation set is also the "\emph{best-effort}" validation data set based on community attribute. 

Table 3 shows the experimental results on three snapshots selected at the same time each year from 2016 to 2018. We use the same validation sets for different methods. It can be clearly seen from the Table 3 that the three traditional inference algorithms have shown good performance and generalization in terms of inference accuracy, which is generally above 93\%, even up to 97\%. By comparison, the \emph{Accuracy} of those feature-based methods is relatively low, which is greatly affected by the used features, dataset and models. For this simple task, our AS-GCN is comparable with the traditional inference methods, with its accuracy above 94\%.

\subsection{Multi-Classification}
Classification of multiple relationships for Internet is more difficult than binary classification based the analysis in Section~\ref{Sec4}. We argue that considering more kinds of links could help us describe the Internet topology more precisely, so as to better understand its evolution mechanism.

Since S2S and X2X relationship only occupy a small part of the entire Internet, in order to prevent the poor training effect caused by the severe imbalance of samples in different types, we randomly sample P2P and P2C relationships to make the four types balanced. To this end, we make $P2P:P2C:S2S:X2X\approx{1:1:1:1}$, and we also divide the four types of AS relationships into separate training, validation and test sets, with a ratio of 6:2:2 for multi-classification experiments. Meanwhile, it is difficult to reflect the performance of the model if the \emph{Accuracy} is used as the only evaluation metric. We thus also consider \emph{Precision}, that how many of the samples predicted to be positive are truly positive samples, and \emph{Recall}, which indicates how many positive examples in the original sample were predicted correctly.

Multi-classification results are shown in TABLE~\ref{Tab4}, where we can see that AS-GCN behaves surprising well in inferring the four kinds of relationships, with accuracy, precision, and recall all exceed 91\%, outperforming RF, Xgboost and LightGBM, although it has been shown that it is quite difficult to distinguish S2S from P2C and X2X from P2P just based on several typical features. Note that the traditional inference methods, such as AS-Rank, ProbLink and TopoScope, are specially designed for binary-classification, which cannot be used for multi-classification task, and thus they are not listed here. Such results validate again that our AS-GCN is more general and can be used to effectively recognize different kinds of relationships between ASes.

\begin{table*}\caption{The \emph{Accuracy} of AS link inference methods for binary classification experiment.}
\renewcommand\arraystretch{1.3}
\centering
\renewcommand{\multirowsetup}{\centering}
\begin{tabular}{c|c c c|c c c|c}
\bottomrule[2pt]
\multirow{3}{1.8cm}{Date} & \multicolumn{3}{c|}{Traditional inference methods} & \multicolumn{3}{c|}{Feature-based methods} & \multicolumn{1}{c}{\multirow{3}{1.8cm}{AS-GCN}} \\ 
\cline{2-7}
& \multirow{2}{1.8cm}{AS-Rank} & \multirow{2}{1.8cm}{ProbLink} & \multirow{2}{1.8cm}{TopoScope} & \multirow{2}{1.5cm}{\begin{tabular}[c]{@{}c@{}}RF\end{tabular}} & \multirow{2}{1.3cm}{Xgboost} & \multirow{2}{1.8cm}{LightGBM} & \multicolumn{1}{c}{} \\
&&&&&&& \multicolumn{1}{c}{} \\ 
\hline
2016 & 96.75 & \textbf{96.80} & 96.44 & 93.14 & 94.92 & 91.74 & 95.90 \\ 

2017 & 94.19 & 95.08 & 93.77 & 82.04 & 91.10 & 88.90 & \textbf{95.93} \\ 

2018 & \textbf{97.37} & 96.55 & 96.21 & 79.46 & 92.65 & 86.18 & 94.66 \\ 
\toprule [2pt]
\end{tabular}
\end{table*}

\begin{table*}[]
\caption{The \emph{Accuracy}, \emph{Precision} and \emph{Recall} of the three feature-based methods and our AS-GCN model for multi-classification experiment.}
\renewcommand\arraystretch{1.3}
\centering
\begin{tabular}{c c|c|c|p{1cm}<{\centering} p{1cm}<{\centering} p{1cm}<{\centering} p{1cm}<{\centering}|p{1cm}<{\centering} p{1cm}<{\centering} p{1cm}<{\centering} p{1cm}<{\centering}}
\bottomrule[2pt]
\multirow{2}{0.5cm}{Date}     & \multirow{2}{0.5cm}{Links} & \multirow{2}{1cm}{Methods} & \multirow{2}{1cm}{\begin{tabular}[c]{@{}c@{}}Accuracy\\ (all)\end{tabular}} & \multicolumn{4}{c|}{Precision} & \multicolumn{4}{c}{Recall} \\ 
\cline{5-12} 
&&&& P2P    & P2C   & S2S   & X2X   & P2P   & P2C   & S2S  & X2X  \\ 
\hline
\multirow{4}{0.5cm}{2016} & \multirow{4}{*}{11093}      
& RF & 94.71 & 94.37 & \textbf{99.79} & 88.99 & \textbf{95.19} & 97.20 & 97.20 & \textbf{94.29} & 90.49 \\ 
&& Xgboost      & 94.66 & \textbf{94.87} & 98.99 & 89.34 & 94.86 & 96.20 & 97.80 & 93.81 & 91.04 \\ 
&& LightGBM     & 94.26 & 94.43 & 98.58 & 91.63 & 92.24 & 95.00 & 97.00 & 91.19 & 90.49 \\ 
&& AS-GCN       & \textbf{95.32} & 94.37 & 99.60 & \textbf{92.36} & 94.60 & \textbf{97.20} & \textbf{98.80} & 92.14 & \textbf{92.87} \\ 
\hline
\multirow{4}{0.5cm}{2017} & \multirow{4}{*}{10962}      
& RF & 92.62 & 90.17 & 99.17 & 86.68 & 94.66 & 97.20 & 95.20 & 92.76 & 85.85 \\ 
&& Xgboost      & 93.28 & 92.69 & 98.37 & 87.21 & \textbf{94.72} & 96.40 & 96.40 & \textbf{94.12} & 86.78 \\ 
&& LightGBM     & 93.33 & 92.34 & 98.39 & 87.83 & 94.41 & 96.40 & 97.60 & 91.40 & 88.08 \\ 
&& AS-GCN       & \textbf{95.54} & \textbf{95.85} & \textbf{99.40} & \textbf{93.76} & 93.28 & \textbf{97.20} & \textbf{99.20} & 91.03 & \textbf{94.42} \\ 
\hline
\multirow{4}{0.5cm}{2018} & \multirow{4}{*}{13227}      
& RF & 92.98 & 91.80 & 93.96 & 91.23 & 94.08 & 94.00 & 96.40 & 88.30 & 93.26 \\ 
&& Xgboost      & 93.96 & 93.69 & 95.45 & 91.35 & \textbf{94.77} & 95.00 & 96.40 & 89.62 & 94.57 \\ 
&& LightGBM     & 92.69 & 91.49 & 96.78 & 91.99 & 91.57 & 94.60 & 96.20 & 84.53 & 94.46 \\ 
&& AS-GCN       & \textbf{95.38} & \textbf{95.42} & \textbf{99.00} & \textbf{93.26} & 94.60 & \textbf{95.80} & \textbf{99.40} & \textbf{92.09} & \textbf{94.94} \\ 
\toprule[2pt]
\end{tabular}
\label{Tab4}
\end{table*}

\begin{figure*}[htbp]
\centering
{
    \begin{minipage}[b]{\linewidth}
        \centering
        \includegraphics[scale=0.43]{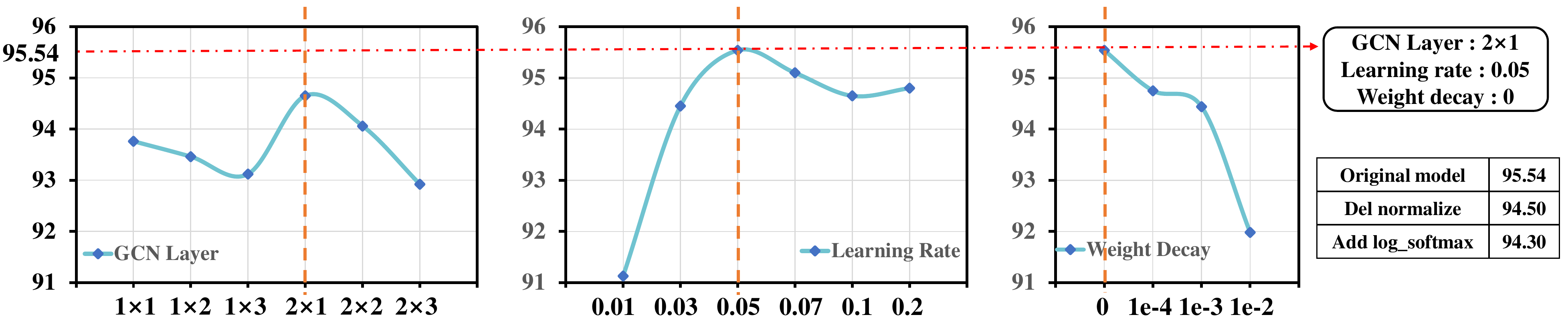}
    \end{minipage}
}
\caption{Parameter sensitivity analysis for AS-GCN model.}
\label{Fig7}
\end{figure*}

\subsection{Parameter and Feature Analysis}
In this subsection, we firstly conduct some sensitivity analysis of hyper-parameters on the 04/01/2017 dataset. We analyze how different choices of the hyper-parameters may affect the performance of our AS-GCN. Second, we analyze the importance of seven features in detail by controlling other variables. Finally, we make a visual analysis for the multi-classification results of AS links.

\begin{figure*}[tbh]
\centering
{
    \begin{minipage}[b]{\linewidth}
        \centering
        \includegraphics[scale=0.9]{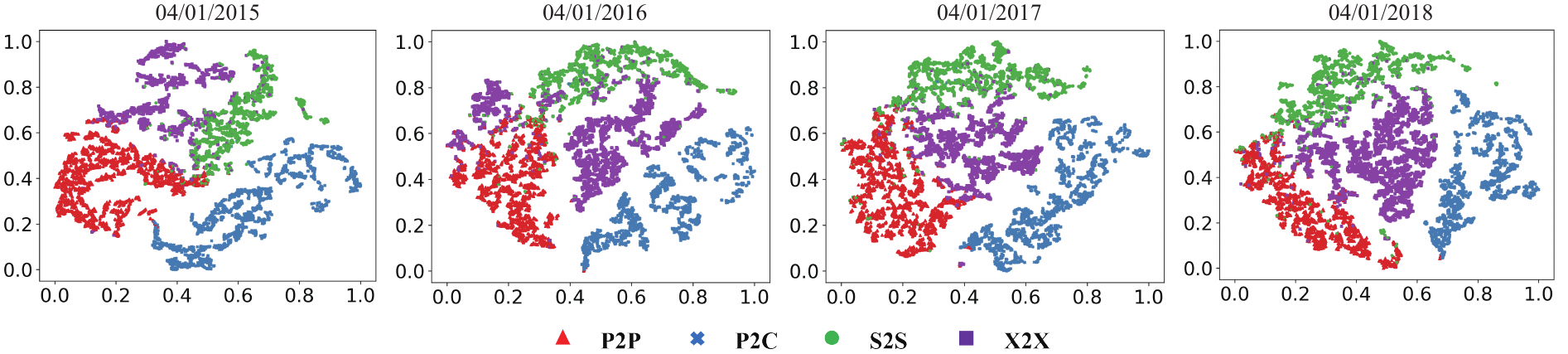}
    \end{minipage}
}
\caption{Visualization of multi-classification obtained by AS-GCN using t-SNE algorithm.}
\label{Fig9}
\end{figure*}

Fig.~\ref{Fig7} gives the parameter optimization process of AS-GCN on 04/01/2017 dataset for the multi-classification. The main parameters considered here are GCN structure, learning rate and weight decay. We obtain the optimal parameter in each round of experiments and then use it as a fixed value before evaluating the next parameter. In Fig.~\ref{Fig7}, the optimized parameters are obtained as follows: The setting of AS-GCN structure is $2 \times 1$. The learning rate is 0.05. And the model does not have obvious overfitting phenomenon, so the weight decay parameter is set to 0. Simultaneously, through experimental analysis, the final output of each GCN block needs to be added the normalization operation after activation function \emph{ReLU}. We share the same setting of parameters on different experimental datasets.

\begin{table}[]
\caption{Feature importance analysis}
\renewcommand\arraystretch{1.3}
\begin{tabular}{ p{3.7cm}<{\centering}|p{1cm}<{\centering}|p{1.3cm}<{\centering}|p{1cm}<{\centering} } 
\bottomrule[2pt]
Features removed        & Acc. & \begin{tabular}[c]{@{}c@{}}Importance\\ score\end{tabular} & \begin{tabular}[c]{@{}c@{}}Average\\ ranking\end{tabular} \\ \hline
Common neighbor ratio     & 91.45     & 15.60 & \textbf{3.25} \\ 
\hline
Transit degree            & 91.51     & 15.37 & 7.75 \\ 
\hline
Distance to VP mean       & 92.72     & 10.76 & 4.75 \\ 
\hline
Distance to clique        & 92.88     & 10.14 & \textbf{2} \\ 
\hline
Degree                    & 92.96     & 9.84 & 3.75 \\ 
\hline
Hierarchy                 & 93.01     & 9.65 & 8.25 \\ 
\hline
AS type                   & 93.29     & 8.58 & 7 \\ 
\hline
Distance to VP min        & 93.30     & 8.54 & 6 \\ 
\hline
Assign VP                 & 94.03     & 5.76 & 5.75 \\ 
\hline
Distance to VP max        & 94.03     & 5.76 & 6.5 \\ 
\toprule[2pt]
\end{tabular}
\label{Tab5}
\end{table}

\begin{figure}[h]
\centering
{
    \begin{minipage}[b]{\linewidth}
        \centering
        \includegraphics[scale=0.34]{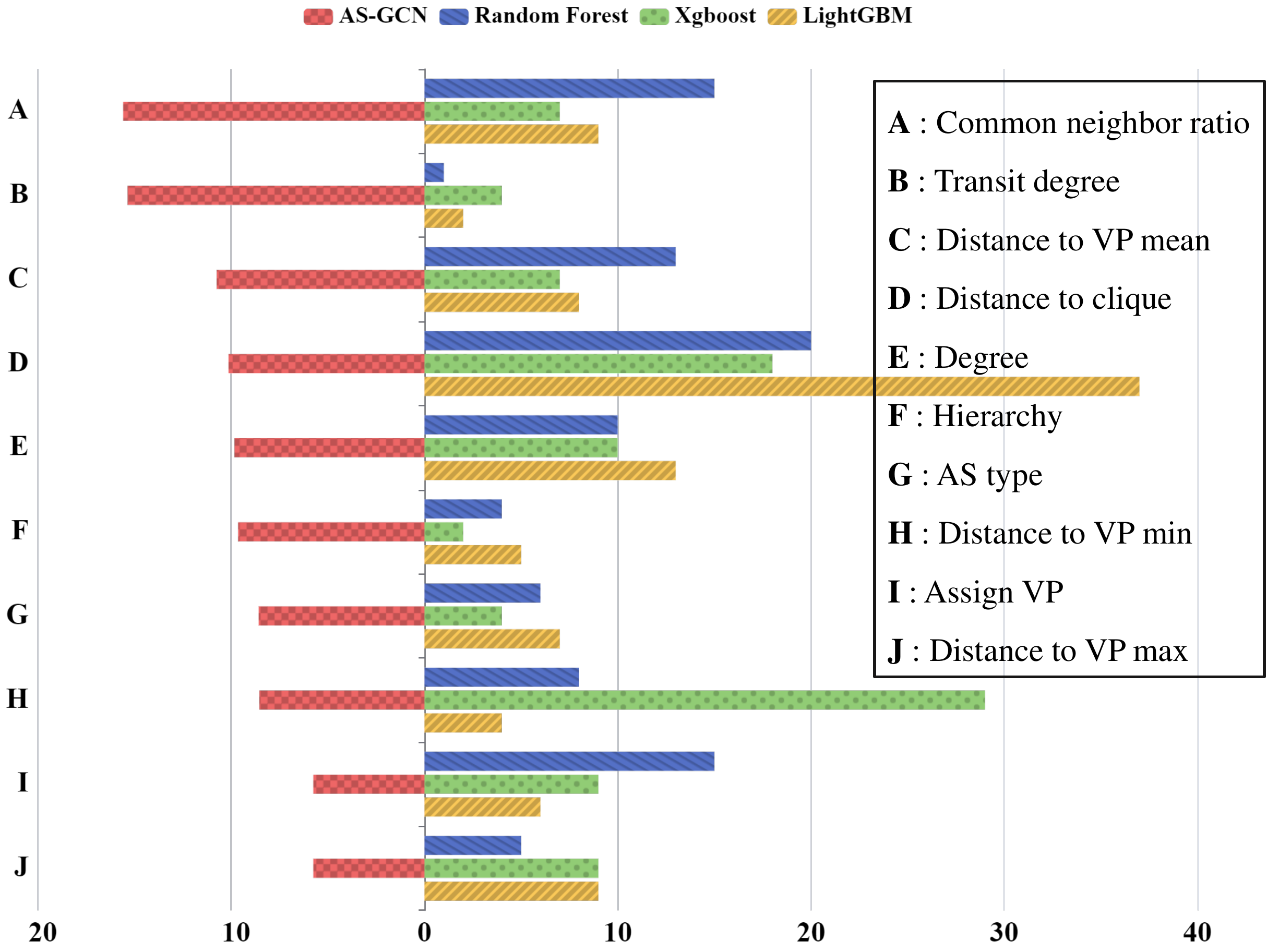}
    \end{minipage}
}
\caption{Feature importance analysis on AS-GCN and its comparison methods}
\label{Fig8}
\end{figure}

Fig.~\ref{Fig8} and TABLE~\ref{Tab5} show the features importance, which is a method of scoring and sorting the input features of a classification model, and reveals the relative importance of each feature when making classification. We evaluate each features of the AS-GCN model using \emph{Accuracy} by removing only one specified feature from the all at once to forms the left side of the histogram in Fig.~\ref{Fig8}. In addition, the feature importance evaluation results construct under other three comparison algorithms on the right side of the figure. For AS-GCN, the importance score (denote as $S$) can be formulated as follows:
\begin{equation}
S_i = \frac{| \operatorname{Acc}_{org} - \operatorname{Acc}_{i} |}{\sum_{j=1}^{d} | \operatorname{Acc}_{org} - \operatorname{Acc}_{j} |} \times 100 \%
\label{import}
\end{equation}
where $\operatorname{Acc}_{org}$ represents the \emph{Accuracy} of the original model, and $\operatorname{Acc}_{i}$ represents the \emph{Accuracy} by removing the $i^{\rm th}$ dimension feature. From Fig~\ref{Fig8} and TABLE~\ref{Tab5}, we can see that \emph{common neighbor ratio} first proposed by us is the most important feature in AS-GCN, whereas \emph{distance to clique} and \emph{distance to VP} is relatively important feature in feature-based methods. In addition to the above, we also average the ranking of importance scores obtained by the four methods. The average ranking is listed in TABLE~\ref{Tab5}.
Obviously, \emph{distance to clique} and \emph{common neighbor ratio} are manifested as the strong importance features in each algorithm. On the contrary, \emph{transit degree} as the exclusive feature of AS-level network topology, it does not show excellent performance.

Finally, we use a dimensionality reduction algorithm \emph{t-SNE}~\cite{van2008visualizing} to project the high-dimensional vector output by AS-GCN into the two-dimensional space for intuitive visualization. As show in Fig.~\ref{Fig9}, we can clearly observe the degree of aggregation and discrimination between the four types of AS relationship. This proves the effectiveness of our proposed AS-GCN framework once again.

\section{Conclusion and Future Work}
With the in-depth study of previous inference algorithms and the importance of relevant AS features, we find that S2S and X2X relationships are difficult to infer by existing algorithms. Thus, except those typical AS features that have been proposed, we also introduce two new AS features from the graph perspective to distinguish these two types of AS links. Meanwhile, we propose a GCN-based model AS-GCN, achieving outstanding results compared with many baselines on the highly reliable datasets we construct.

In the future, we still face inevitable challenges. On the basis of this paper, we will focus more on the inference of complex relationships (i.e., hybrid and partial transit relationships) with machine learning methods. The dataset problems have always been a particularly critical part of machine learning methods. The labels of our datasets also come from inference algorithms in nature, so we can’t guarantee that they are 100\% correct. Like previous researchers, we are eager for a standard dataset to better evaluate and further improve our inference methods. On the other hand, although we have proposed several new features, such as \emph{AS type} and \emph{common neighbor ratio}, there still lack features that have a strong distinction to make multi-classification judgments. Therefore, in the follow-up study of various types of business relationships, it is also very important to find more representative features of each kind of relationships. In general, the effective use of our methods in practice is our ultimate goal.


%



\section*{Acknowledgment}
The authors would like to thank all the members of the IVSN Research Group, Zhejiang University of Technology for the valuable discussions about the ideas and technical details presented in this paper.

\ifCLASSOPTIONcaptionsoff
  \newpage
\fi



%

\bibliographystyle{IEEEtran}
\bibliography{myref}
%








\end{document}